\input phyzzx
\hoffset=0.375in

\def\kpc{\rm kpc}
\def\max{{\rm max}}
\def\au{\rm AU}

\def\kms{\rm km\,s^{-1}}

\def\eps{{\epsilon}}

\font\bigfont=cmr17
\centerline{\bigfont Non-Acceleration of Sgr A*:}
\smallskip
\centerline{\bigfont Implications for Galactic Structure}
\bigskip
\centerline{{\bf Andrew Gould}\foot{Alfred P.\ Sloan Foundation Fellow}
and \bf Solange V.\ Ram\'\i rez}
\bigskip
\centerline{Dept of Astronomy, Ohio State University, Columbus, OH 43210}
\smallskip
\centerline{E-mail:  gould,solange@astronomy.ohio-state.edu}
\bigskip
\singlespace
\centerline{\bf ABSTRACT}

We show that observations by Backer and collaborators over the past two 
decades constrain the time derivative of the proper motion of Sgr A* to be 
$<0.14\,\rm mas\,yr^{-2}$.  Using this result and a preliminary measurement
by Eckart \& Genzel of $\sigma\sim 500\,\kms$ for the velocity dispersion of 
the star cluster within $0.\hskip-2pt ''2$ of Sgr A*, we derive the following 
implications.   First, if the nuclear star cluster is dominated by a massive 
black hole, then either Sgr A* is that black hole or it orbits the black hole 
with a radius $\lsim 3\,$AU.  Second, even if the star cluster does not 
contain a massive black hole, Sgr A* is constrained to move slower than 
$20\,\kms$ ($1\,\sigma$) relative to the center of mass of the cluster.  The 
Galactocentric distance is therefore $R_0=7.5\pm 0.7\,$kpc, independent of the 
nature of Sgr A*.  These error bars could be substantially reduced by future 
observations.  If they are, it will also be possible to probe the motion of 
the nuclear star cluster relative to the center of mass of the Galaxy at the 
$\sim 4\,\kms$ level.

Subject Headings:  astrometry -- Galaxy: center, fundamental parameters
\endpage
\bigskip

\chapter{Introduction}

	Backer \& Sramek (1987) and Backer (1996) have measured the 
proper motion of the compact non-thermal radio source Sgr A* and find values
that are in reasonable agreement with those expected from the reflex motion of 
the Sun, assuming an LSR rotation speed $v_{\rm LSR}=220\,\kms$ and a 
galactocentric distance $R_0=7.5\,$kpc.  The $1\,\sigma$ error bars 
($\sim 0.15\,\rm mas\ yr^{-1}$) correspond to $\sim 5\,\kms$ at a distance
of 7.5 kpc.

	This agreement has been used as the basis for two related
arguments.  Backer (1996) reasons that since the source is moving at most
very slowly relative to the proper motion of the Galactic center as
predicted from ``known'' Galactic parameters, it must be a black hole
of at least $100\,M_\odot$.  On the other hand, Reid (1993) reasons that since 
the source is very likely to be at rest with respect to the center of mass
of the Galaxy, one can use its proper motion to measure $v_{\rm LSR}/R_0$, and
so (to the extent that $v_{\rm LSR}$ is considered known) constrain $R_0$.
Each argument is important and interesting, but clearly both cannot be used
together.

	Here we show that upper limits on the {\it time derivative} of the
proper motion of Sgr A* imply that it is very nearly at rest with respect to
nuclear star cluster at the Galactic center.  One may therefore use this
proper motion to draw conclusions about Galactic structure independent of
any assumptions about the nature of the source.

\chapter{Observational Data}

	Comparing the results of Backer \& Sramek (1987),
$$(\mu_l,\mu_b) = (-5.95\pm 0.70,+0.43\pm 0.50)\,\rm mas\,yr^{-1},
\qquad (1987),\eqn\backprev$$
and Backer (1996), 
$$(\mu_l,\mu_b) = (-6.55\pm 0.17,-0.48\pm 0.12)\,\rm mas\,yr^{-1},
\qquad (1996),\eqn\backlast$$
it is clear that the proper motion of Sgr A* did not change much over a decade.
It is difficult to give a precise upper limit to the time derivative of the
proper motion because the underlying data
have not been published.  For purposes of this paper, we estimate the
upper limit by combining equations \backprev\ and \backlast\ with the data 
points shown in Figure 1 of Backer \& Sramek (1987) and find
$$\biggl[\biggl({d^2 l\over d t^2}\biggr)^2 +
\biggl({d^2 b\over d t^2}\biggr)^2\biggr]^{1/2}
< 0.14\,\rm mas\,yr^{-2},\eqn\pmdot$$
at the $1\,\sigma$ level,
corresponding to a limit on the physical transverse acceleration $a_\perp$ of
$$a_\perp < a_\max = 5\,\kms\,\rm yr^{-1}\sim  0.025\,a_\oplus,\eqn\amax$$
where we have for simplicity of exposition adopted $R_0=7.5\,\kpc$, and where
$a_\oplus$ is the acceleration of the Earth.  We believe that this estimate
is conservative, but in any event we indicate below how the results depend
on $a_\max$.

	From the work of Menten et al.\ (1997) the position on infrared
images corresponding to the radio position of Sgr A* is now known to an
accuracy of $0.\hskip-2pt ''03$.  Eckart \& Genzel (1997) have measured the
proper motions of stars in the infrared within $2''$ ($\sim 0.1\,$pc) of this
position.  In general, they find that the velocity dispersion rises toward
the center in a way that is consistent with a central black hole with mass
$M_* = 2.45\times 10^6\,M_\odot$.  In particular, for the measurement 
at the innermost point at $r_* = 2000\,\au$, they find a velocity dispersion
$\sigma\sim 500\,\kms$, i.e., still consistent with a central black hole of
mass $M_*$ at the center of the subregion of radius $r_*$,
$$M_* = 2.4\times 10^6\,M_\odot,\qquad r_* = 2000\,\au.\eqn\mstarrstar$$
Eckart \& Genzel (1997) regard the measurement at this last point
as still preliminary.
For purposes of this paper, we assume that a mass $M_*$ is contained within
a radius $r_*$, so that the magnitude of the acceleration at the boundary of 
this region is
$$ a(r_*) = {M_*/M_\odot\over (r_*/\au)^2}a_\oplus\sim 0.6\,a_\oplus.
\eqn\aofrstar$$
As more data are acquired, it will be possible to refine the estimates
of Eckart \& Genzel (1997).  The results presented here can then be rescaled.

	Equations \amax\ and \aofrstar\ reveal the basic result that we will
exploit in this paper: the ratio, $\eps$, of the upper limit for the 
transverse  acceleration of Sgr A* to the characteristic acceleration of the 
system in which it is embedded is very small,
$$\eps \equiv {a_\max\over a(r_*)} \sim 0.04\ .\eqn\epsdef$$
To understand the implications of this result,
we consider two limiting cases: first where the mass $M_*$ is dominated by
a single point mass (a black hole), and second where $M_*$ is distributed
uniformly throughout the region inside $r_*$.  We demonstrate that in either
case, Sgr A* is moving at most very slowly with respect to the center of
mass of the star cluster in which it is embedded.

\chapter{Kepler Potential}

	Suppose that the region within $r_*$ is dominated by a massive
black hole.  Then there are two possibilities: either Sgr A* 
{\it is} that black hole
or it is orbiting in the potential of the black hole.  If the first is true,
our case is already proved, so we restrict consideration to the second.

	We designate the position of Sgr A* relative to the black hole by
$(r,\theta,\phi)$ where $\theta$ is the angle Sun--black hole--Sgr A*.  Then,
$$\sin\theta = {a_\perp\over a(r)} < {a_\max\over a(r)} = \eps
\biggl({r\over r_*}\biggr)^2.\eqn\sinthetalimit$$
The prior probability for such a fortuitous geometry at any given instant 
is extremely small, less than $(3/10)\eps^2 \sim 5\times 10^{-4}$ for a 
monotonically decreasing density profile.  Even if Sgr A* happened to 
lie sufficiently close to the line of sight to the black hole at the
beginning of the observations, it would move out of this zone within the 
$T\sim 10$ years of observations unless it were on a highly radial orbit, with 
its transverse speed $v_\perp$ constrained by 
$v_\perp < v_\max=\eps r^3/r_*^2 T\sim 40\,\kms\,(r/r_*)^3$.  This further
reduces the prior probability by a factor 
$(v_\max/\sigma)^2/2$ to a net probability of $\lsim 10^{-7}$.  
That is, this scenario is essentially ruled out.  

	Hence, if there is large black hole in the center of the nuclear
star cluster, then Sgr A* must be it.  The one potential loophole is that 
Sgr A* might be physically associated with the black hole and orbit it
with a period much shorter than the 
frequency of observations, $\sim (450\,\rm day)^{-1}$ (Backer \& Sramek 1987).
The physical association would evade the above probability argument, and the
short
period would imply that Sgr A* would orbit many times between observations 
and therefore would not show any secular acceleration.  However, for an
orbital radius $r\lsim 150\,$AU (corresponding to a period $<450\,\rm days$), 
the typical displacement between observations would be 
$\sim r/R_0\sim 20\,{\rm mas}\, (r/150\,\au)$.  The actual displacements from
uniform motion are $\lsim 5\,$mas (see Fig.\ 1 from Backer \& Sramek 1987),
implying that Sgr A* has an orbital radius $r\lsim 40\,\au$, and therefore a 
speed $v\gsim 7000\,\kms$.
Such large velocities are all but excluded by the observations of 
Rogers et al.\ (1994) who put an upper limit of $3.3\,\au$ on the size of 
Sgr A* using observations taken on 1994 April 2 and 1994 April 4.  The authors
note that the observations were phased on NRAO 530 because the signal from
Sgr A* was too weak.  If Sgr A* had a transverse motion greater than 
$7000\,\kms$, it would have moved more than 8 AU between the two sets of
observations, which would have
undermined the phasing and prevented Rogers et al.\ (1994) from setting an
upper limit of 3.3 AU on the size (see Fig.\ 1 of Rogers et al.\ 1994).
Only if $r<3\,\au$ (or if the orbital inclination and phase were particularly 
unfavorable) could this conclusion be avoided.
However, even if Sgr A* were in such an orbit (as opposed to being the black
hole) its observed proper motion would 
still be equal to the proper motion of the black hole, since the size of the
orbit would be a small fraction of the $\gsim 700\,$AU that Sgr A* has been
observed to move relative to the Sun over the lifetime of the observations.

	It is possible to test directly the hypothesis that Sgr A* is in
a small orbit, $r\lsim 3\,\au$, by looking for time variability of the flux 
due to the Doppler effect.  The fractional amplitude of the flux oscillations 
would be $f\sim(1+p)v\sin i/c$, where $p\sim 0.33$ is the slope of power law
($S_\nu\propto \nu^p$, Mezger 1996), $i$ is the orbital inclination, and
$v$ is the orbital velocity.  Thus $f\sim 0.12(r/3\,\au)^{-1/2}\sin i$ with
period $\sim 1.2\,{\rm days}\,(r/3\,\au)^{3/2}$.

\chapter{Harmonic Oscillator Potential}

	Next, we suppose that the mass within $r_*$ is not concentrated at
a point, but rather is a distributed throughout the region, perhaps in the
form of stars or possibly other objects.  Most likely, the density profile
would be monotonically decreasing, but for simplicity and to focus on an
extreme case, we consider a uniform distribution.  We note that there are
many potential problems for a star cluster of this density because of the
shortness of the relaxation time.  However, the mass need not be in the form 
of stars, but could be in much lighter particles such as weakly interacting 
massive particles (WIMPs).  Alternatively, the problems associated with dense 
star clusters might be avoided by some effect that has so far escaped 
recognition.  Since our 
purpose is to develop completely general arguments, 
we do not make {\it any} assumption about the nature of the material within 
$r_*$, other than that it has total mass $M_*$.

	A uniform distribution gives rise to a harmonic oscillator potential,
so $a(r) = (r/r_*)a(r_*)$.  Thus, the analog of equation \sinthetalimit\ is
$$\rho = r\sin\theta = {a_\perp\over a(r)}r < {a_\max\over a(r_*)}r_* = 
\eps r_*,\eqn\sinthetalimitho$$
where $\rho$ is the projected separation of Sgr A* from the center of the
cluster.  The prior probability for this is 
$(3/2)\eps^2\sim 2\times 10^{-3}$.  As in
the case of the Kepler potential, Sgr A* would have to be on a nearly
radial orbit.  Including
both effects, the prior probability is 
$(3/4)\eps^4(r_*/\sigma T)^2\sim 7\times 10^{-6}$.  Again, the scenario
is ruled out.

	The one exception to this argument would be if Sgr A* were
{\it gravitationally confined} to be near the center.  Then it would 
not be at a random position in the cluster, and the previous probability
argument would fail.  In order to be sufficiently confined to satisfy 
equation \sinthetalimitho, $\rho<\eps r_*$, its characteristic speed would
be constrained by 
$$v\lsim \eps\sigma\sim 20\,\kms.\eqn\vlimit$$
Thus even in this case,
the proper motion of Sgr A* would be the same as that of the cluster center
of mass to within $\sim 9\%$.  

	If the density profile fell monotonically (giving rise to a potential
intermediate between Kepler and harmonic-oscillator) the arguments presented
in this section would still hold but with greater force:  the fraction of
phase space satisfying the constraint \amax\ would be even smaller than
$7\times 10^{-6}$, and the maximum of velocity of an object gravitationally
confined to a region that did satisfy the constraint would be even less than
$20\,\kms$.

	We note in passing that by equipartition, 
the minimum mass of Sgr A* required for it
to be gravitationally confined as described above is
$M_{{\rm Sgr }\ {\rm A*}} > \eps^{-2}M\sim 600\,M$,
where $M$ is the characteristic mass of the objects in the cluster.

\chapter{Implications for Galactic Structure}

	Unfortunately, at the present time no hard and fast conclusions
can be drawn from the lack of observed acceleration of Sgr A*.  The
arguments given above rest crucially on the Eckart \& Genzel's (1997)
observation of a high dispersion, $\sigma\sim 500\,\kms$, within 
$0.\hskip-2pt ''2$ of Sgr A*.  Since those authors regard their result
as preliminary, any conclusions that are drawn from these observations
must have the same caveat.  Nevertheless, their preliminary
result is quite plausible and
could well be confirmed within a few years.  We therefore begin by
assuming that it will be confirmed and investigate the consequences.

	Since equation \vlimit\ limits the motion of Sgr A* relative to
the Galactic center to $<20\,\kms\sim 0.6\rm\,mas\,yr^{-1}$ at the 
$1\,\sigma$ level, one can apply the approach of Reid (1993) to constrain 
$R_0$ but without making any assumptions about the nature of Sgr A*.  Of
course, the price of relaxing these assumptions is the additional 
uncertainty in the motion of the Galactic center of $20\,\kms$.  
To be specific, we adopt an estimate with $1\,\sigma$ error
of $v_{\rm LSR} = 220\pm 10\,\kms$.  We note in passing that the small size
of this error bar rests critically on the assumption that the rotation curve
of the Galaxy, like that of similar external galaxies, is flat.  With this
assumption, measurement of the redshifts of tangent points interior to 
the Sun lead to an estimate very close to $v_{\rm LSR}\sim 220\,\kms$
(Brand \& Blitz 1992).  If this assumption is dropped, the error estimate
increases by several fold.  We assume that the Sun is moving at $12\,\kms$
relative to the LSR, or $232\pm 10\,\kms$ relative to the Galactic frame.
We make use of Backer's (1996) measurement
and $1\,\sigma$ error, $\mu_l=-6.55\pm 0.17\,\rm mas\,yr^{-1}$ and find,
$$R_0 = 7.5\pm 0.7\,\kpc\qquad ({\rm provisional}).\eqn\rzeroest$$

	Even when the Eckart \& Genzel (1997) measurement is confirmed, 
some of the most interesting information about Galactic structure will
remain inaccessible due to the relatively weak upper limit (eq.\ \pmdot) to the
time derivative of the proper motion.  This limit can probably be significantly
improved simply by fitting existing data to a second order polynomial.  In
any event,  continued observations at a uniform rate and with uniform
quality would yield a rapid improvement in the precision of this quantity,
$\propto T^{-5/2}$, where $T$ is the total duration of the observations.
Hence, we also investigate what can be learned if the upper limit \pmdot\ can
be significantly improved.

	Of course, the uncertainty in equation \rzeroest\ would be 
reduced.  However, it would also be possible to probe an entirely different
question: whether the nuclear star cluster at the Galactic center is at
rest with respect to the center of mass of the Galaxy.  At a distance
of $7.5\,\kpc$, Backer's (1996) proper-motion measurement in the $b$
direction, $-0.48\pm 0.12\rm\,mas\,yr^{-2}$ ($1\,\sigma$), translates into
$-17\pm 4\,\kms$.  The Sun's motion relative to the LSR is $7\,\kms$ and is
extremely well measured, with an uncertainty of $\ll 1\,\kms$ (Mihalas \&
Binney 1981).  Hence there is a net motion of $-10\pm 4\,\kms$ that remains
unexplained.  At the present time, it is not possible to draw any conclusion
about this residual for three reasons.  First, the effect itself is 
detected at only the $2.5\,\sigma$ level and so could be just a statistical
fluctuation.  Second, the LSR may be moving relative to
the Galactic frame because of a warp in the disk or some other effect.
It is possible to directly test this hypothesis.  Although the data available 
to date
are inconclusive, significant improvements could be made in the
future (see below).  Third, the observed deviation is completely consistent
with the constraint \vlimit\ on the motion of Sgr A* relative to the cluster.
This limit is directly proportional to $\eps$ and so to $a_\max$ (see eq.\
\epsdef).  If the upper limit on $a_\max$ could be reduced by a factor 
$\sim 10$, this would remove relative motion of Sgr A* as a potential cause
of this effect.  The same continued proper motion measurements would improve
the statistical error on the proper motion in the $b$ direction.  Thus there
would remain two potential causes, motion of the LSR and motion of the
central star cluster (e.g.\ Miller \& Smith 1992), 
both of which are interesting possibilities.

	The motion of the LSR can be investigated by finding the mean motion
of the Sun relative to stars in the Galactic halo.  The velocity ellipsoids
of several populations have been measured including 162 RR Lyraes 
(Layden et al.\ 1996; Popowski \& Gould 1997), 887 non-kinematically
selected metal-poor field stars (Beers \&
Sommer-Larsen 1995), and 1352 high proper-motion stars 
(Casertano, Ratnatunga, \& Bahcall 1990).
However, only the authors of the RR Lyrae studies report on (or fit for)
the mean $z$ motion of their sample, $\VEV{W}$.  
Popowski \& Gould (1997) find  
$\VEV{W}= -13\,\pm 8\,\kms$, which is consistent at the $1\,\sigma$ level 
with both the LSR value of $-7\,\kms$ and the Sgr A* value of $-17\,\kms$.
It is straightforward to analyze the sample of Beers \& Sommer-Larson (1995),
since the data are publicly available and since the selection criteria
are non-kinematic.  We conduct a joint analysis of the Beers \& 
Sommer-Larson (1995) stars and Layden et al.\ (1996) stars.  To obtain a
homogeneous sample, we restrict the former to stars within 3 kpc of the Sun 
and delete the variables (which are mostly RR Lyraes), and we restrict both 
samples to stars with $\rm [Fe/H]<-1.5$.  This leaves a total of
724 Beers \& Sommer-Larson (1995) stars with radial velocities and 106
Layden et al.\ (1996) stars with both radial velocities and proper motions.
Using the method of Popowski \& Gould (1997) we find
$$\VEV{W} = -6\pm 5\,\kms,\eqn\vzeq$$
very close to the LSR value of $\sim -7\,\kms$.  
It is truly unfortunate that Casertano et al.\ (1990) did not
fit for $\VEV{W}$ because at this point it would not be at all easy to
reimplement the beautiful technique they devised to remove even unrecognized
selection biases in their samples.  If their samples were reanalyzed, 
however, we estimate that
the error in the determination of $\VEV{W}$
would be $\sim 5\,\kms$.  Thus, by combining the Casertano et al.\ (1990)
sample with the results summarized in equation \vzeq\
based on the
RR Lyraes and the non-kinematically selected stars, one could reduce the 
uncertainty to $\sim 3.5\,\kms$.  Hence, if the discrepancy persists between 
the $z$ motion of the LSR and that of Sgr A*, it should be possible to decide 
which of them is actually moving relative to the Galactic center of mass.

{\bf Acknowledgements}:  
We thank G.\ Newsom for a careful reading of the manuscript.
This work was supported in part by grant AST 94-20746 from the NSF.

\Ref\backer{Backer, D.\ C.\ 1996, in Unsolved Problems of the Milky Way,
eds., L.\ Blitz and P.\ Teuben (IAU)}
\Ref\backer{Backer, D.\ C., \& Sramek, R., A.\ 1987, in AIP Conf.\ Proc.\
ed., D.\ L.\ Backer, p.\ 163, (New York: AIP)}
\Ref\beers{Beers, T.\ C., \& Sommer-Larsen, J.\ 1995, ApJS, 96, 175}
\Ref\bb{Brand, J.\ \& Blitz, L.\ 1992, A\&A, 275, 67}
\Ref\crb{Casertano, S., Ratnatunga, K.\ U., \& Bahcall, J.\ N.\ 1990, ApJ,
357, 435}
\Ref\eckart{Eckart, A., \& Genzel, R.\ 1997, MNRAS, 284, 576}
\Ref\lay{Layden, A.\ C., Hanson, R.\ B., Hawley, S.\ L., Klemola, A.\ R., 
\& Hanley, C.\ J.\ 1996, AJ, 112, 2110}
\Ref\ms{Miller R.\ H., \& Smith, B.\ F.\ 1992, ApJ, 393, 508}
\Ref\menten{Menten, K.\ M., Reid, M.\ J., Eckart, A., \& Genzel, R.\ 1997,
ApJ, 475, L111}
\Ref\mes{Mezger P.\ G.\ 1996, in The Galactic Center, R.\ Gredel ed.
ASP Conference Series, Vol 102, p 380} 
\Ref\MB{Mihalas, D.\ \& Binney, J.\ 1981, Galactic Astronomy, (San Francisco:
Freeman)}
\Ref\pop{Popowski, P., \& Gould, A.\ 1997, ApJ, submitted}
\Ref\reid{Reid, M., J.\ 1993, ARA\&A, 31, 345}
\Ref\rogers{Rogers, A., E., E., et al.\ 1994, ApJ, 434, L59}

\endpage
\refout
\end